\documentclass[aoas,MSNbibl,nameyear,dvips]{arximspdf}
\usepackage{dcolumn}
\usepackage{graphicx}

\doi{10.1214/12-AOAS559} 
\volume{6}
\issue{4}
\pubyear{2012}
\firstpage{1531}
\lastpage{1551}

\makeatletter
\newcolumntype{d}[1]{D{.}{.}{#1}}

\makeatother

\begin{document}
\begin{frontmatter}

\title{Phenotypic evolution studied by layered stochastic differential
equations\thanksref{T1}}
\thankstext{T1}{Supported by CEES at the University of
Oslo, the Research Council of Norway (RCN Project 197823) and
by the Royal Swedish Academy of Sciences through funding by the
Knut and Alice Wallenberg Foundation (KAW 2009.0287).
This project used DSDP/ODP samples provided by the Integrated
Ocean Drilling Program (IODP). IODP is sponsored by the U.S.
National Science Foundation (NSF) and participating countries
under the management of Joint Oceanographic Institutions (JOI), Inc.}
\runtitle{Layered SDEs}

\begin{aug}
\author[A]{\fnms{Trond} \snm{Reitan}\corref{}\ead[label=e1]{trondr@bio.uio.no}},
\author[B]{\fnms{Tore} \snm{Schweder}\ead[label=e2]{tore.schweder@econ.uio.no}}
\and
\author[C]{\fnms{Jorijntje} \snm{Henderiks}\ead[label=e3]{jorijntje.henderiks@geo.uu.se}}

\address[A]{T. Reitan\\
CEES\\
Department of Biology\\
University of Oslo\\
P.O. Box 1053 Blindern\\
N-0316 Oslo\\
Norway\\
\printead{e1}}
\address[B]{T. Schweder\\
Department of Economics\\
University of Oslo\\
P.O. Box 1095 Blindern\\
N-0316 Oslo\\
Norway\\
\printead{e2}}
\address[C]{J. Henderiks\\
CEES\\
Department of Biology\\
University of Oslo\\
P.O. Box 1053 Blindern\\
N-0316 Oslo\\
Norway\\
and\\
Department of Earth Sciences\\
Uppsala University\\
Villav{\"{a}}gen 16\\
SE-75 236 Uppsala\\
Sweden\\
\printead{e3}}

\runauthor{T. Reitan, T. Schweder and J. Henderiks}

\affiliation{University of Oslo, University of Oslo, and University of
Oslo and~Uppsala~University}
\end{aug}

\received{\smonth{3} \syear{2012}}
\revised{\smonth{4} \syear{2012}}

%
\begin{abstract}
Time series of cell size evolution in unicellular marine algae
(division Haptophyta; \textit{Coccolithus} lineage), covering 57
million years,
are studied by a system of linear stochastic differential
equations of hierarchical structure. The data consists of
size measurements of fossilized calcite platelets (coccoliths)
that cover the living cell, found in deep-sea sediment cores
from six sites in the world oceans and dated to irregular points in time.
To accommodate biological theory of populations tracking their
fitness optima, and to allow potentially interpretable correlations
in time and space, the model framework allows for an upper layer of partially
observed site-specific population means, a layer of site-specific
theoretical fitness optima and a bottom layer representing
environmental and ecological processes. While the modeled process
has many components, it is Gaussian and analytically tractable.
A total of 710 model specifications within this framework are
compared and inference is drawn with respect to model structure,
evolutionary speed and the effect of global temperature.
\end{abstract}

%

\end{frontmatter}

\section{Introduction}\label{sec1}
Biological populations evolve with respect to the distribution of organism
size and other phenotypic traits by differential fitness. How a
phenotypic character like body size evolves over time and what
environmental factors influence phenotypic change are fundamental
questions of biology and paleontology. Some key issues include the
following: Is
the average size fluctuating around a fixed or a changing fitness
optimum [\citet{estesarnold2007}]? How fast does the population mean
track the optimum? What is the inertia of the fitness optimum, and
is it tracking a deeper process representing environmental conditions?
If there is a deeper process, what qualities does it have? Are
population-specific optima related, for example, by influence from
an unobservable underlying global climatic process? Alternatively,
are the phenotypic means at different sites correlated directly?
Is the optimum responding to global temperature?

Several recent studies have highlighted a covariance
between the body size evolution of marine organisms and
global temperature. The long-term cooling trend over the
past 65 million years appears to be matched by a
macroevolutionary size decrease in marine phytoplankton
[\citet{finkel2005}] while marine zooplankton
and benthos show size increases [see \citet{schmidt2004}, \citet{hunt2010}].
With the exception of \citet{hunt2010}, which only used
random walks as a time series model, these studies did,
however, not take time dependency into
account when testing this hypothesis. Without
consideration of the internal dynamics of two trended
time series, the null hypothesis of no interaction
between the series may be wrongfully rejected.
Here, we study the evolution of body size in marine algae by means of
layered stochastic differential equation (SDE) models.

\begin{figure}

\includegraphics{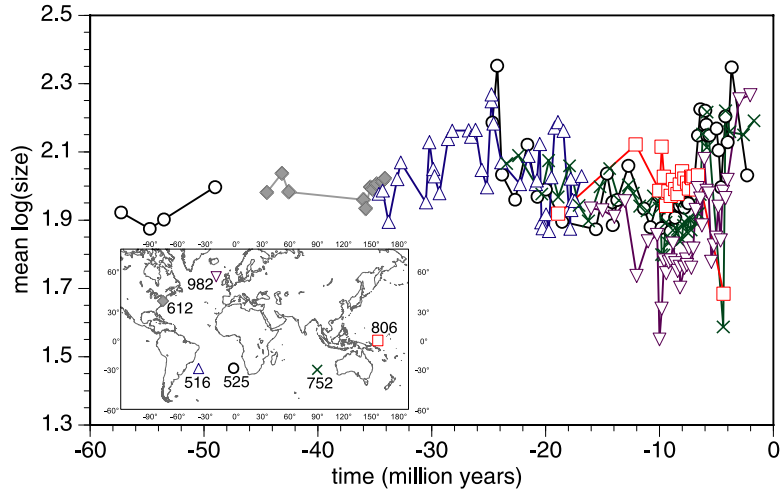}

\caption{DSDP and ODP site locations (inset map) and site-specific mean
logarithmic size values spanning the last 57 \textup{My}. One symbol
is shown for each of the 205 observations of mean log
coccolith size in the time series plot.}
\label{mapandmeasurements}
\end{figure}

In unicellular algae, cell size and its geometry largely determine
the transport rates of dissolved components (e.g., CO${}_2$/O${}_2$, nutrients)
into and out of the cell, which is fundamental to photosynthesis and
growth rates. We study the evolution in average log-phenotype of
calcifying microalgae belonging to the \textit{Coccolithus}
Schwartz 1894 lineage. The \textit{Coccolithus} lineage
(division Haptophyta) has
extant species in today's oceans and a well-documented fossil
record dating back to the early Paleocene [see \citet{haqlohmann1976}].
We study coccolith size, which is measured by the largest
diameter in the elliptically shaped calcite platelets (coccoliths)
that cover the unicellular organism.
Coccolith size is used as a proxy for cell size [see \citet{henderiks2008}].
A total of 205 deep-sea sediment samples, taken from 6 different
sites in the Atlantic, Indian and Pacific oceans, offer a final
data set of 19,899 individual size measurements distributed over
the last 57 million years (My) (Figure~\ref{mapandmeasurements}).
To current standards of palaeontology and evolutionary biology, these
data are
unusually extensive.

%

The population of algae behind sampled coccoliths from one site is assumed
to track its fitness optimum by natural selection. Fitness
(expected number of reproducing offspring) is distributed over the
individuals of the population, with the optimum varying over time
according to physical and ecological conditions. The fitness optima
of different populations may be spatially correlated, due to common
environmental conditions, in addition to being temporally correlated
within a population. What affects fitness with respect to coccolith
size is unknown. Global mean ocean surface temperature might be a
contributor, and a measured temperature indicator series [\citet{zachos2001}]
is tested in the model framework as potentially driving the
fitness optimum processes along with other underlying but
unmeasured processes.

In this study, we will work on the logarithmic scale, as untransformed
coccolith sizes cannot be normally distributed. Mean logarithmic
coccolith size
for each population is seen as a stochastic process pulled toward
an optimal state, which itself is seen as a stochastic
process subject to pull from additional underlying
processes. These underlying processes might be population specific
and possibly correlated across populations. The model framework
thus allows for three hierarchical layers of processes,
each layer having one process for each geographical location.
In the building of our framework, the three layers are the
population means, population fitness optima and underlying
environmental processes affecting fitness such as an observed
global temperature indicator series and also unmeasured
environmental variables.

As the data are irregularly distributed in time,
a stochastic time series framework that can handle continuous time
is called for. Linear SDE models constitute a parsimonious
framework for such modeling. Vector processes governed by
linear SDEs, as defined in Section~\ref{SDEs}, are suitable for
hierarchical models with variables ordered by the flow of causality
as determined by the coefficients
in the equations; see \citet{schweder2012}.
Linear vectorial SDEs are analytically tractable, having an
explicit likelihood.

A collection of 710 variants of models within this framework
were fitted to the sample means at the six locations.
These model variants are compared and properties of the
evolutionary process are discussed by way of commonalities
in models which are singled out by various statistical criteria.
We use both Bayesian and classic methods.

\citet{lande1976} was the first to suggest that
the population mean of a phenotypic trait,
for example, logarithmic coccolith size, might evolve like an
Ornstein--Uhlenbeck (OU) process. This was taken
further by \citet{estesarnold2007}.
The Lande model [\citet{lande1976}] for the evolution
of the mean phenotype, such as
logarithmic size, is governed by the one-dimensional linear SDE,
\begin{equation}
dX(t)=-\alpha\bigl(X(t)-\mu_0 \bigr)\,dt+\sigma \,dW(t), \label{OU}
\end{equation}
where $dW(t)$ is white noise and $t$ is time. This is an OU
process when $\alpha>0$. The mean phenotype
$X$ is pulled toward the level $\mu_0$ with a force proportional to
its displacement $X(t)-\mu_0$ both when having been pushed by randomness
above or below its long-term level $\mu_0$. This level is the
fitness optimum in \citet{lande1976}. We will use the term pull for
the parameter $\alpha$ and diffusion for $\sigma$. Phenotypic
evolution is also modeled by OU processes in
\citet{hansen1997}, though in a phylogenetic rather than
time series perspective, and so do \citet{hansen2008}, where the
level process $\mu(t)$ is defined to be an underlying random walk.
A random walk model, which can be seen as a limiting case of
equation (\ref{OU}) with $\alpha=0$, has been studied by \citet{hunt2006}
and \citet{hunt2008}.
Random walks have been in use for a long time, as a proposed
null hypothesis in evolutionary models; see, for instance,
\citet{raup1977}.
SDE models have been around for years, and have been
used in biology [see \citet{allen2003}, Chapter 8]
and physics [see \citet{schuss1980}, Chapter 2].

Our aim is to develop a class of models that is capable of
describing the data reasonably well, that allows biological
interpretations and that is statistically feasible.
These models might also be useful elsewhere,
especially for time series with irregular temporal resolution.

In Section~\ref{SDEs} we will first introduce the linear SDE
framework tailored for our application.
We will then briefly discuss issues of causality and hierarchy.
Model selection and inference is described in Section~\ref{inference}.
These methods are then applied first to artificial data in Section
\ref{sim} and then to the coccolith data in
Section~\ref{coccolith}. Some concluding
remarks are offered in Section~\ref{conclusions}.

\section{The coccolith data}
Microfossils (coccoliths) were measured in a total of 205
sediment samples obtained from Deep Sea Drilling Project (DSDP) and
Ocean Drilling Program (ODP) deep-sea sediment cores, taken at six
sites in the Atlantic, Indian and Pacific oceans, altogether spanning
about 57 My; see Figure~\ref{mapandmeasurements}.

In each sample, 1 to 400 individual coccoliths were measured
on slides using polarized light microscopy
[see \citet{henderiks2006}, \citet{henderiks2008}], resulting in a
mean sample size of $19\mbox{,}899/205=97$. The data was transformed
from original size in $\mu m$ to the logarithm of that, before
means and variances were calculated. The age of each sample
is estimated from biostratigraphic data calibrated to the
geological time scale of \citet{candekent1995}.

While the coccoliths within a single sample may have been
formed as much as one thousand years apart, we consider
them simultaneously formed on our geological time scale.
The sampling process is assumed random, from the
historical population through deposition fossilization,
drilling and extraction from the drill cores.

The stationary distribution of mean log size is normal in
our model. Similarly, any
lack of normality in the samples should give little cause for
concern when modeling phenotypic means by sample means, due
to the central limit theorem.
In the tradition following \citet{lande1976}, we study the
evolution of the population mean of the phenotype in question,
which in our case is the logarithmic body size.
Population variance and other population processes and
parameters are of potential interest, but for our study these
are nuisance parameters. To keep the complexity of our model
within bounds dictated by the data, the population variances
are estimated outside the SDE framework and smoothed by a simple
GAM analysis so as to reduce the variance uncertainty for
samples with a low number of measurements.

We assume the sample means to be normally distributed with known
variances; see equation (\ref{obs}) in the next section. The 19,899 separate
measurements are thus summarized by the 205 sample means. The
assumption of normality and known variances is not strictly true, but
not far off, and it allows the likelihood to be computed.

To diagnostically check the assumption that the sample means of log
coccolith size are normally distributed, skewness and kurtosis in
the distributions of the sample means were estimated by bootstrapping
for the 192 samples with more than 4 observations, and tested. The
$p$-value plots [see \citet{schwederspjoetvoll1982}] show only small
departures from normality, with some sample means having a bit of
positive skewness, but with no trace of kurtosis. See the
supplementary information for more on the normality of the data, the
treatment of the measurement variances and other data issues.

\section{Layered linear SDE models}
\label{SDEs}
\subsection{The modeling framework}
If we concentrate on a single population, the idea
is to connect measurements from different times together
using a continuous time series model. Thus, the irregular
time intervals between observations will not pose a problem.

As described in the \hyperref[sec1]{Introduction}, we aim at describing
the phenotypic mean as a process responding to another
continuous time process which is hidden, namely, the size at
optimal fitness, which may again respond to environmental
changes, partially or entirely unmeasured. Thus, the equation
describing the dynamics in
phenotypic mean (layer one) will not only contain the phenotypic mean,
but also a hidden underlying process
representing optimal fitness (layer two). Layers are in general processes
defined so that one layer can respond to the
current state in another layer, which is then said to be below it.
The layer may then in its turn affect the
layer above it, unless it is the topmost layer.

In our framework, layer two can again be affected by a third layer,
understood as the unmeasured environmental conditions.
The second layer might also be forced by an external global temperature
indicator time series on this layer, as measured by \citet{zachos2001}.

In an SDE, the change in continuous time series from one time point
to another time point infinitesimally further along is modeled
by a transformation of the previous state plus some
normally distributed noise. When this transformation is linear and
the observational noise is assumed normal and not conditioned on
the state, we call this a (vectorial) linear SDE. When one
has a linear SDE system and the state at the starting point of the
process is normally distributed, then both the marginal distribution
and the conditional distribution of the state of a later time point
is also normally distributed. When also the sampling errors are
normal, a~likelihood can be calculated. These conditions
provide the modeling framework for this study.

A sample mean $Y_t$ for time $t$
is thus considered a noisy representative of
the state of the topmost layer $X_1(t)$, namely, the phenotypic
population mean at the same time point. So
\begin{equation}
Y_t \sim N\bigl(X_1(t),s_t^2/n_t\bigr),
\label{obs}
\end{equation}
where $s_t^2$ is the sample variance and $n_t$ is the sample size.
The sampling errors $Y_t-X_1(t)$ are assumed independent.
In the language of hidden Markov models, equation (\ref{obs}) will be
the observation equation while the system equation will
be described in the next subsections.
Since the state process is a normally distributed linear hidden
Markov chain and the observations are normal and independent given the
state process, the Kalman filter can be applied for calculating the
likelihood; see the supplementary material [\citet{supp}].

As there are six geographic sites, instead of a single process
per layer, we have six components in each layer.
To allow for possible instantaneous commonalities in a layer, there
might be instantaneous correlations across sites in the layer.
We allow only correlations between processes belonging to the same layer.

Collected over the 6
sites, the state of the system is situated in an 18-dimensional vector
space and evolves according to a linear hierarchical system of SDEs.

\subsection{Linear SDE basics and examples}
\label{sde}
The OU process, described in equation (\ref{OU}), is a
simple linear SDE process. It is
stationary when $\alpha>0$. The stationary
distribution is normal with expectation $E(X(t))=\mu_0$
and standard deviation $\eta\equiv \operatorname{sd}(X(t))=\sigma/\sqrt{2\alpha}$.
This can be verified by equation (\ref{cov}) later in the text.
The stationary correlation is $\operatorname{corr}(X(t_{1}),X(t_{2}))=
\mathrm{e}^{-\alpha(t_{2}-t_{1})}=\mathrm{e}^{-(t_{2}-t_{1})/\Delta t}$,
$t_{1}<t_{2}$, where $\Delta t\equiv1/\alpha$ gives
a characteristic correlation time for the process, the time for
the correlation to drop to $1/\mathrm{e}$.
Such alternate parametrizations, using, for instance, $\Delta t$ and
$\eta$ instead of the pull $\alpha$ and the diffusion $\sigma$, can make
the interpretation of results easier, as well as help in the
elicitation of Bayesian priors. Thus, with three parameters, one can
parsimoniously describe a continuous function of time that is a
stationary process.

When the level term $\mu_0$ in equation (\ref{OU}) is replaced
by a process $X_2(t)$ of the OU type, one gets a coupled linear SDE:
\begin{eqnarray}\label{OU2}
dX_1(t) &=& -\alpha_1\bigl (X_1(t)-X_2(t)\bigr)\,dt + \sigma_1
\,dW_1(t),
\nonumber
\\[-8pt]
\\[-8pt]
\nonumber
dX_2(t) &=& -\alpha_2\bigl (X_2(t)- \mu_0\bigr)\,dt + \sigma_2 \,dW_2(t),
\end{eqnarray}
where
$X_1$ is the topmost, partially measured process layer. It is driven
by $X_2$, which is a hidden process. Since $X_2$ is unaffected by $X_1$,
it is an OU process on its own. $X_1$ is, however, dynamically
affected by $X_2$ and has a different time correlation function
from that of an OU process, as we will show later.
This causal structure is described by
$X_2\rightarrow X_1$. When this causality takes the
form found in equation (\ref{OU2}), $X_1$ is said to be tracking $X_2$.

The joint process of $X_1$ and $X_2$ stacked in a vector is a
special case of the following vectorial linear SDE,
\begin{equation}
dX(t)=\bigl(m(t)+AX(t)\bigr)\,dt+\Sigma \,dW(t), \label{SDEbig}
\end{equation}
where the state vector $X(t)$ and $m(t)$ are $p$-dimensional,
$A$ is
called the pull matrix and is of dimension $p\times p$,
$m(t)$ is regarded a deterministic process,
$\Sigma$ is a $p\times q$ dimensional diffusion matrix and
$W$ is a $q$-dimensional Wiener process. The structure of
direct causal relations between components of $X$
is determined by the nonzero elements in the pull
matrix $A$. If $A_{i,j} \neq0$, then component $X_j$ directly
affects $X_i$, and there is an arrow from $X_j$ to $X_i$
in the causality graph.
The notion of causality used here is equivalent to that of
\citet{granger1969} in Markov process models; see
\citet{schweder2012}.
Since the contributions in equation (\ref{SDEbig}) are normal and
linear, the distribution will be normal and thus
characterized by the expectation and covariance:
\begin{eqnarray} \label{cov}
EX(t) &=& V^{-1} \mathrm{e}^{\Lambda(t-t_0)} V X(t_0)+
V^{-1} \int_{t_0}^t \mathrm{e}^{\Lambda(t-u)} V m(u) \,du,
\nonumber
\\[-8pt]
\\[-8pt]
\nonumber
\operatorname{cov}(X(v),X(t)) &=&
V^{-1} \biggl[ \int_{t_0}^v \mathrm{e}^{\Lambda(v-u)} V \Sigma(u)\Sigma
(u)^{\prime} V^{\prime}
\mathrm{e}^{\Lambda(t-u)} \,du \biggr] (V^{-1})^{\prime}
\end{eqnarray}
for $v\leq t$, where $V$ is the left eigenvector matrix and
$\Lambda$ is the diagonal matrix of eigenvalues of $A$, so that
$A=V^{-1}\Lambda V$ conditioned on the existence of such a representation.
A layered causal structure is achieved by specifying which
of the elements in $A$ are nonzero and by restricting the
diffusion matrix to a blocked structure. See the supplementary
material [\citet{supp}] and \citet{schweder2012} for more on this.

Since the covariance structure of $X(t)$ governed by
equation (\ref{SDEbig}) is available in a closed form in equation
(\ref{cov}),
the covariance of the topmost component of the process of
equation (\ref{OU2}) can be found to be
\begin{eqnarray}\label{auto-covar2}
\operatorname{cov} ( X_1 ( 0) , X_1 ( t) ) =
\frac{\sigma_1^2}{2\alpha_1}\mathrm{e}^{-\alpha_1 t}+
\frac{\sigma_2^2 \alpha_1^2} {\alpha_1^2-\alpha_2^2}
\biggl( \frac{1}{2\alpha_2}
\mathrm{e}^{-\alpha_2 t}-\frac{1}{2\alpha_1}
\mathrm{e}^{-\alpha_1 t}\biggr) ,
\end{eqnarray}
provided $\alpha_1 \neq\alpha_2$ are both positive.
The covariance structure here is different from the one-layered
OU process described in equation (\ref{OU}). This
indicates that data generated from the top layer of a two-layered
process will be detectably different from that of the one-layered case.

The model we initially made for the coccolith size data had
three layers, phenotypic mean, optimum and climate.
We assumed the stochastic contributions in the
second layer to be identical over the sites, while the stochastic
contributions in the two other layers were uncorrelated over the sites.
We have $k=6$ processes indexed by $j$ representing each of the
sites in each of the 3 layers, denoted by $X_{1,j}$ in the top layer, $X_{2,j}$
in the middle layer and $X_{3,j}$ at the bottom. The flow of
causality is
\begin{equation}
X_{3,j} \rightarrow X_{2,j} \rightarrow X_{1,j}.
\label{origgraph}
\end{equation}

So, for a given site $j \in\lbrace1, \ldots, 6 \rbrace$, we have
the following SDE system:
\begin{eqnarray}\label{OriginalModel}
dX_{1,j}(t)&=&-\alpha_1\bigl(X_{1,j}(t)-X_{2,j}(t)\bigr)\,dt + \sigma_1
\,dW_{1,j}(t),
\nonumber\\
dX_{2,j}(t)&=&-\alpha_2\bigl(X_{2,j}(t)-X_{3,j}(t)-\beta T(t)\bigr)\,dt + \sigma_2
\,dW_2(t),
\\
dX_{3,j}(t)&=&-\alpha_3\bigl(X_{3,j}(t)-\mu_0\bigr) \,dt + \sigma_3 \,dW_{3,j}(t),\nonumber
\end{eqnarray}
where the $W$ processes are independent Wiener processes. The regressor
term $\beta T(t)$ defines how the second layer responds
to an exogenous process $T(t)$, which in our application
will be a global temperature indicator series, as measured by
\citet{zachos2001}. Thus, $m(t)$ in equation (\ref{SDEbig}) will
have a contribution $\alpha_3 \mu_0$ on the third layer
and a contribution $\alpha_2 \beta T(t)$ on the second.
The second layer thus tracks a linear combination of
the third layer and the external time series $T(t)$.
While there are 18 processes in this particular model, it has
only 8 parameters, namely, $\theta=
(\alpha_1,\alpha_2,\alpha_3,\sigma_1,\sigma_2,\sigma_3,\beta,\mu_0)$.
The processes are regional (though inter-regional instantaneous
correlation is introduced in layer two) but the parameters are global
and thus so
is the nature of the dynamics.
The covariance function on the top (phenotypic) layer for this model is
shown in the supplementary material [\citet{supp}].

\subsection{Pull identifiability}
\label{identifiability}
It is shown in the supplementary material [\citet{supp}] that in any multi-layered
process having only one pull parameter per layer, the pulls can
be reshuffled so that if there are $l$ layers, then
$\alpha_l<\alpha_{l-1}<\cdots<\alpha_2<\alpha_1$.
It thus initially seemed natural to restrict the inference
outcomes so that this is the case, which will be called the
pull identification restriction. However, in the case of multiple
sites, requiring, for example, that
$\alpha_3<\alpha_{2,s}<\alpha_1$ for all sites $s$ will exclude
some covariance structures. When one analyzes data using the model
framework without imposing the pull identification restriction,
one does, however, need to keep in mind that a reshuffling of the pull
parameters is possible and may in some cases even be necessary.
The alternative is either to impose the restriction only
on one specific site or to impose them on all and so risk overlooking
valid solutions. As we wished to avoid singling out a specific site
and did not want to risk being overtly restrictive, we did not
enforce the pull identification restriction. The identification issue
should, however, be kept in mind when interpreting the results. In the
supplementary material [\citet{supp}] we present results also under the pull the
identification restrictions.

When initially testing the simulation study, we used the pull
identification restriction. The simulation analysis suggested an error
in our treatment of the combination of prior distribution and
the identification restriction, which resulted in Bayesian model
selection support for over-complex models. For
one-layered models the prior distribution for the pull parameter
remains undisturbed. However, when imposing the
pull identification restriction on our prior distribution for
multi-layered models, the marginal distribution of each pull
parameter became narrower. Thus, if the data suggests a pull parameter
within this narrower interval, this is better predicted by
the multi-layered model which is then unfairly supported in a Bayesian
model likelihood
comparison. We thus corrected our prior distribution so that the
marginal distribution for at least one common pull parameter remains
the same for models having different numbers of layers. See the supplementary
material [\citet{supp}] for more on this issue and for the simulation results after
correcting the
prior distribution.

\section{Likelihood-based inference methodologies}
\label{inference}
\subsection{Single model inference}
\label{singelmodel}
A process governed by equation (\ref{SDEbig}) is Gaussian
with known mean and covariance structure. The measurement errors
then add extra variance (equal to the sample variance divided by the
sample size), which is also regarded as known.
The likelihood of observations of the process affected by independent
normal measurement errors is thus available in closed form.
The likelihood function is, however, often multimodal since it
usually is a nonlinear restriction of the covariance matrix
of the observations; see, for example, \citet{sundberg2010}.
Our experience is also that the coccolith data lead to
multimodal likelihood functions for the models we consider.
For maximum likelihood (ML) estimation a shotgun approach with
a hill-climbing algorithm from some 50 starting points widely
distributed in
parameter space seemed to work reasonably well, as long as the
number of parameters were kept low. Stable and efficient ML
results were, however, only gotten when the hill-climbing algorithm
was initialized using posterior Bayesian Markov chain Monte Carlo
(MCMC) samples.

In a Bayesian setting using MCMC sampling, the multimodality
issue seems to have been dealt with and stable, reproducible
results were produced.

Due to the nonlinear parameters
and the relatively low number of measurements, the Hessian at
the likelihood maximum cannot be expected to provide
reliable measures of estimation uncertainty.
Bootstrapping might work, but is computationally
expensive. Bayesian MCMC samples seem, however, to work
and do yield approximate credibility intervals and scatter
plots showing the dependency in the inference on different
parameter combinations.

For the application to the fossil coccolith data, to be
discussed next, we thus used Bayesian methods both for model
choice and for parameter estimation, though we also used
Akaike's Information Criterion (AIC) in the model selection and
included ML estimates in the final presentation of models.

Further description of the Kalman filter, numerical ML optimization,
Bayesian MCMC techniques for single model inference,
numerical problems and computational efficiency are
provided in the supplementary material [\citet{supp}].

\subsection{Model variants}

Our model frame for the coccolith data is an 18-dimensional
version of equation (\ref{SDEbig}) with 3 hierarchical layers
with 6 components each. The diffusion matrix $\Sigma$ is block
diagonal accordingly, with $\Sigma_i$ as the block for layer $i$.
We also consider similar models having one or two
hierarchical layers. For each layer $i$, we consider the
following variants:
\begin{longlist}[(1)]
\item[(1)] Regionality: It could be that one or several parameters
(expectation, layer-specific pull or diffusion) are regional, that is,
different for different sites. Regionality is, however, only allowed
in one of the three layers in a model variant. For more on this,
see the supplementary material [\citet{supp}].

\item[(2)] Determinism: A layer will be deterministic if it
receives no stochastic contributions, that is, if
$\Sigma_i=0$. Note that the pull
and characteristic time may still be finite, so that
the layer performs a deterministic filtering of the
underlying stochastic layer.

\item[(3)] Random walk: This is only possible for the lowest
layer (interpreted as the climatic layer in this
application). With a random walk layer, the process is
nonstationary. Approximate random walk is achieved by setting the
pull very low. Ideally, the pull should be set to zero, but for
practical reasons, we chose to implement $\alpha_i=0.001$,
which means $\Delta t_i=1$~Gy.

\item[(4)] Regional correlation: The components in a layer might be
instantaneously correlated. We allow only one correlation
coefficient, $\Sigma_{i,j_1,j_2}=\rho_i\sigma_{i,j_1}\sigma_{i,j_2}$
for any two sites $j_1 \ne j_2$. With $\rho_i=1$, the instantaneous
stochasticity [associated with $dW(t)$ in equation (\ref{SDEbig})]
in the layer collapses to one dimension, as
for the second layer in equation (\ref{OriginalModel}).
\end{longlist}

By crossing model variants and by varying restrictions on the
parameters, we end up with 710 different models within our model framework.
Without restrictions to only one layer with respect to regionality,
the number of model variants to consider
would much exceed 710, which already is a big number.
Methods for exploring properties of the evolution of
\textit{Coccolithus} by confronting these models with the data in a
Bayesian way is discussed next.

\subsection{Model and property comparison}
\label{modelprob}

In order to calculate Bayesian model probabilities, one
needs the marginal data likelihood,
\begin{equation}
f(D|M) \equiv\int f(D|\theta,M) \pi(\theta|M) \,d\theta,
\label{marg}
\end{equation}
where $D$ is the data, $M$ is the model, $\theta$
is the vector of parameters and $\pi$ is the Bayesian prior
distribution of the parameters. Since this is conditioned on
the~model, we will call it the Bayesian model likelihood (BML).
From this and the prior model probability $\operatorname{Pr}(M)=1/710$, the
posterior model probability $\operatorname{Pr}(M|D)$ is calculated by Bayes'
theorem. Since the prior model probabilities are assumed equal,
$\operatorname{Pr}(M|D)$ is proportional to the marginal likelihood, $f(D|M)$.
Comparing models by their posterior probabilities, one should note
that the posterior probabilities cannot be interpreted in absolute
terms. The problem here is that there might be many models that are very
similar to each other, and the posterior probability of this type of model
will be diluted by the model multiplicity.

The Bayes factor of two competing models is defined by how much the
relative probability for the two models change from the prior to
the posterior case, $B_{1,2}=(\operatorname{Pr}(M_1|D)/\operatorname{Pr}(M_2/D))/
(\operatorname{Pr}(M_1)/\operatorname{Pr}(M_2))=
f(D|M_1)/\break f(D|M_2)$. Thus, it is equal to the marginal likelihood ratio.
The Bayes factor describes the amount of evidence in the
data for model 1 versus model 2.

In equation (\ref{marg}) the parameter $\theta$ is the same for
all model specifications. These models differ, however, with
respect to restrictions imposed on the parameter, and the
prior distribution is modified accordingly. We use a very
wide prior distribution; see below. Due to the complexity
of the prior and nonlinearities in the model, the marginal
data likelihood is not analytically available despite the
likelihood function being Gaussian. The numerical method
we use is an importance sampler described in
the supplementary material [\citet{supp}].

Since we look at many models with different
properties, some such
properties might be evaluated by expressing the different
categories of a given property and
identifying the models within each category.
Looking at such a multitude of models means that many
combinations of properties are tested at once. A model may
then sometimes get support simply by statistical fluctuations.
It will therefore be beneficial to look for a few general
model properties, one at a time, and combine those, when looking
for the best model. In order to
assess what the data indicate concerning different properties,
each category can be given the same property-specific prior probability,
which we will call a weight for the application of studying model
properties. Each model within that property can be given
the same prior weight, so as to better illustrate in which direction
the evidence pulls and how much evidence there is.

If each property was equally probable and each model within each category
also was equally probable, the weights would be Bayesian
probabilities. We are, however, not assuming that this is reasonable.
Instead, we simply use the formalism of Bayesian probability theory, with
probabilities relabeled as weights, in order to explore
how reasonable a given property category is. Note that the ratio
of one weight to another will form the traditional Bayes factor
for a property under the assumption of equal model probabilities
within each property.
If different categorizations are made, this will result in
different prior weights for each model, so it seems more appropriate
to call both the input and the results for weights
rather than probabilities.

A categorization allows for an evaluation of the evidence
for a given category of a property by weighting it with
the number of models it contains. This is done by evaluating the
posterior weights
\begin{equation}
W(C \vert D) = \frac{{1}/{n_C}\sum_{M \in C} f(D \vert M)}
{\sum_{C'}{1}/{n_{C'}}\sum_{M' \in C'} f(D \vert M')} \nonumber
\end{equation}
based on the prior weights $\frac{1}{n_C}$, where $C$ is a category,
$n_C$ is the number of models within the category, and $C'$
runs over the categories of the property.
Thus, this method of comparison compensates for the difference in
number of models with different properties.

\section{Simulations}
\label{sim}
In order to test the model comparison, we made artificial data sets
based on models with one, two and three layers. The models used for
generating the data sets were the best one-, two- and three-layered model
according to BML applied on the original data set while using the ML
estimates for the choice of the parameter set. See the supplementary
material [\citet{supp}] for a description of these three models. Twenty data sets were
then simulated for each of these three cases, with the same amount of data
situated at the same measurement times and having the same measurement
uncertainty as the real data, but where the state processes themselves
were simulated using these models.

The same three models were then used for analyzing each of these
three types of data sets and to see how often different statistical model
selection methods, AIC, BIC and BML, resulted in the right number
of inferred layers. The results are shown in Table~\ref{simmod}.
BIC seems to perform best when a one-layered model has produced
the data, but underperforms for data produced by a three-layered model.
In total, it performs slightly worse than AIC. Fisher's exact test
for contingency tables suggest that there is no significant
difference in performance between AIC and BML for data produced by
a one- and three-layered model, but that BML underperforms in relation
to AIC for data produced by a two-layered model.

%
\begin{table}
\tablewidth=250pt
\caption{Number of correct identifications of a model (as given by the number
of layers) according to different selection schemes, when a total of
$20$ data
sets are simulated; see text}\label{simmod}
\begin{tabular*}{250pt}{@{\extracolsep{\fill}}lccc@{}}
\hline
\textbf{Actual model} & \textbf{AIC} & \textbf{BIC} & \textbf{BML}
\\
\hline
One layer & 16 & 20 & 15 \\
Two layers & 20 & 18 & 12 \\
Three layers & 19 & 15 & 19 \\
\hline
\end{tabular*}
\end{table}

It should, however, be noted that these observations only hold for our
particular choice of one-, two- and three-layered models and is
conditioned on the parameter values behind the simulated data.
We used the maximum likelihood estimates. Since Bayesian methods
allow for property inference, inclusion of prior knowledge and
measures of parameter uncertainty, we will use both AIC and BML in the
analysis. It should also be noted that
according to the results, it is more likely than not that any of these
three methods will identify the correct number of layers, no matter
which of the three models produced the data.

\section{Results and discussion for the \textit{Coccolithus} data}
\label{coccolith}
We ran the model selection exercise without imposing the pull
identification restrictions (Section~\ref{identifiability}).
The model variants selected by BML and AIC as the best violated these
restrictions, and were found substantially better than models satisfying
the restrictions. As a consequence, the analysis shown here is without
pull identification restrictions. Results under these restrictions
are shown in the supplementary material [\citet{supp}].

\subsection{Multi-layered model selection}

Going through all 710 model variants, we opted to do
the model selection both by looking at BML and AIC and by
doing Bayesian property analysis.

%
\begin{table}
\caption{Maximum likelihood (ML) estimates, Bayesian posterior median
(B. median), 95\%~prior~and posterior credibility intervals for the
model considered\break best according to BML.
The units of the characteristic times $\Delta t_{\mathrm
{layer},\mathrm{site}}$ are\break specified
in the table, where the following units are used:
$\mathrm{y}={}$year, $\mathrm{ky}=10^3$ years, $\mathrm{My}=10^6$ years,
$\mathrm{Gy}=10^9$ years. The fixed median of the state
$\mathrm{e}^{\mu_0}$, specified in the original measurement scale,
is given in units of $\mu m=\mathit{micrometers}$. The parameter $\beta$ is the
effect of temperature indicator on the second layer process in units of
$\log(\mathrm{size})/C^{\circ}$. The diffusion parameters $\sigma_{\mathrm
{layer}}$ are
in units of $\log(\mathrm{size})/\mathrm{My}^{1/2}$}\label{bestmodnoid}
\begin{tabular*}{\textwidth}{@{\extracolsep{\fill}}lccccc@{}}
\hline
& \textbf{Parameters} & \textbf{ML} & \textbf{B. median} & \textbf
{Prior 95\%} & \multicolumn{1}{c@{}}{\textbf{Posterior 95\%}} \\
\hline
& $\mathrm{e}^{\mu_0}$ & $7.42$ & $7.43$ &
$0.001$--$1000$ & 7.15--7.69 \\
& $\beta$ & $0.0006$ & $0.0005$ &
$-1$--$1$ & $-0.002$--$0.012$ \\
Upper & $\Delta t_{1}$ & $11$~ky & $16$~ky &
$1$~ky--$1$~Gy & $0.3$~ky--$80$~ky \\
\quad layer & $\sigma_{1}$ & $0.48$ & $0.43$ &
$0.02$--$3.5$ & $0.22$--$2.8$ \\[3pt]
Middle & $\Delta t_{2,525}$ & $130$~ky & $120$~ky &
$1$~ky--$1$~Gy & $1.0$~ky--$0.52$~My \\
\quad layer & $\Delta t_{2,612}$ & $215$~My & $40$~My &
$1$~ky--$1$~Gy & $0.58$~My--$3.2$~Gy \\
& $\Delta t_{2,516}$ & $0.4$~ky & $11$~ky &
$1$~ky--$1$~Gy & $90$~y--$150$~ky \\
& $\Delta t_{2,752}$ & $1.0$~My & $1.0$~My &
$1$~ky--$1$~Gy & $250$~ky--$2.9$~My \\
& $\Delta t_{2,806}$ & $3.7$~Gy & $88$~My &
$1$~ky--$1$~Gy & $2.0$~My--$10$~Gy \\
& $\Delta t_{2,982}$ & $0.4$~ky & $12$~ky &
$1$~ky--$1$~Gy & $100$~y--$140$~ky \\[3pt]
Lower & $\Delta t_{3}$ & $1.2$~My & $1.4$~My &
$1$~ky--$1$~Gy & $0.64$~My--$3.7$~My \\
\quad layer & $\sigma_{3}$ & $0.17$ & $0.16$ &
$0.02$--$3.5$ & $0.11$--$0.24$ \\
& $\rho_3$ & $0.66$ & $0.64$ &
$-0.18$--$0.98$ & $0.29$--$0.85$ \\
\hline
\end{tabular*}
\end{table}

To our knowledge, little information is available
on which to base the prior distribution,
except broad ideas about the time scales involved
in biological processes and general knowledge about
the size and variation of size in single-celled organisms.
We therefore construct our prior by applying independent
normal distributions to each type of parameter
(or rather a reparametrized version so that the new parameters
can be allowed to take values over the entire real line).
See the supplementary information [\citet{supp}] for more concerning
the prior distribution, which was wide but informative.

The best model according to BML (and
second best according to AIC), a three-layered model, is shown in
Table~\ref{bestmodnoid}.
It had a Bayes factor $B=215$ compared to the best one-layered model,
while the AIC difference between the best multi-layered model
according to AIC and
the best one-layered model according to AIC was $\Delta \mathrm{AIC}=-11.2$.
The best model according to AIC was a variant of the best model
according to BML only with an extra inter-regional correlation
found in the first layer.
Since the top layer turned out to have very fast-moving dynamics,
any contributions due to inter-regional correlations
would quickly vanish. Therefore, the structural difference between the
best AIC and best BML model made very little practical difference.
Two fairly different statistical model selection methods thus gave
approximately the same model, having overall the same structure.
There was regional pull without diffusion in the second layer (deterministic
filtering of the lowest layer, with different filtering for
different sites). Inter-regional correlations were inferred to be
on the third layer.

According to Table~\ref{bestmodnoid}, the second layer contains
both sites with greater and smaller pulls than that of the third layer,
which is a result not possible if the pull identification restriction is
imposed. Note that while layers can be switched in a multiple site
setting as well as for single set, there is no way to switch them
so as to make pulls progressively lower for all sites.

Compared to our initial model, there are a couple of changes.
The inter-regional correlation is moved to the lowest layer
and is no longer perfect. The pull parameters on the second layer
are regional rather than global and there is no diffusion on
that layer. Thus, this model has five more parameters than the
initial model.

\subsection{Influence of external data series (global climate)}

After finding the best model according to BML,
we looked at expansions that allowed for an external
data series to influence the second layer though a regression
term, $\beta$.
The 95\% credibility interval for $\beta$ encompassed zero,
indicating cell size to be rather unaffected by the global
temperature indicator. The Bayes factor between the model described
and the same model without the temperature indicator series regression,
$B=f(D \vert M_{\mathrm{no\ temp}})/f(D \vert M_{\mathrm{with\ temp}})$, is about
$662$. The Bayes factor for the regression is sensitive to
the choice of prior for $\beta$, so that a wider 95\% credibility
interval $(-10,10)$, results in the Bayes factor larger than $6000$.
Thus, the strength of the evidence against temperature dependence
is sensitive to the prior, though for both the initial and the
alternative choice of prior distribution for $\beta$, the Bayes
factor indicates no evidence for temperature dependence. With
a 95\% credibility interval encompassing zero and having an
upper limit of about $0.012$ and with a variation in the
temperature indicator series of about $4.7$, the impact, if any, of
the temperature indicator variations on the second layer would with
$95\%$
probability be less that $\operatorname{exp}(4.7 \times0.012) \approx6\%$.
Thus, the global temperature indicator influence on the
coccolith size will be disregarded for the rest of the
analysis.

%
\begin{figure}
\centering
\begin{tabular}{@{}c@{}}

\includegraphics{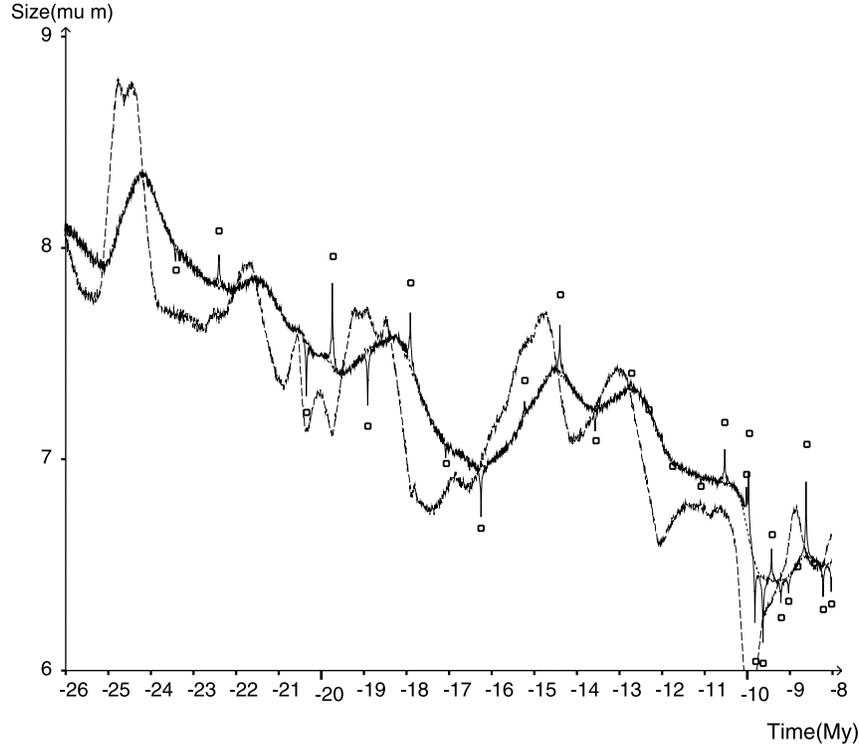}
\\
\footnotesize{(a)}
\end{tabular}
\caption{Process inference and measurements for Site 752 running
in the time interval $-26$~\textup{My} to $-8$~\textup{My}, using the model in Table \protect
\ref{bestmodnoid},
which is considered best according to BML. The inference takes
Bayesian parameter uncertainty into account as well as process
uncertainty given parameter values.
\textup{(a)} Three-layered posterior means.
Solid line: upper layer; short dashed line: middle layer;
long dashed line: lower layer; circles: measurements.
The upper and middle layer may be hard
to separate, except for small time intervals close to measurements.
Note that there are patterns in the lowest layer before the first
measurement belonging to this site. This is because the lowest layer
has correlations to other sites, for which there are measurements before
this time.
\textup{(b)} Upper layer posterior mean, uncertainty and one realization.
Solid line: mean; dashed lines: limits of a 95\% credibility interval;
grey wiggly
line: realization; circles: measurements. Except for the right
side of the graph, where there are many high quality measurements,
the credibility interval is wide.
Note that while the upper layer realization may seem noisy due to
the short-memory stochastic contributions to that layer, it
is also influenced by lower layers with longer memory. This creates
correlations over large time spans, so that the realization can be
over or under the top layer mean for large time spans.}
\label{3layer}
\end{figure}

\subsection{Inference of model properties}
Inference on the process states they themselves can also be performed
in the same framework, using the Kalman smoother method.
An example of this is shown in Figure~\ref{3layer}, where the
three layers belonging to a single site in our data set have been studied
using sampled parameter sets (taken from the posterior MCMC samples)
on the model considered best. Figure~\ref{3layer}(a) indicates that the
top layer
is well adapted to the observations, while the processes of the
layers below have expectation values that are progressively
smoother versions of the layer above. Figure~\ref{3layer}(b) shows
how the uncertainty of the topmost process can be associated
with a fixed maximum span far outside the observations
but becomes progressively smaller the closer in time one is to an
observation. However, the credibility interval does not shrink to
zero, because of observational noise.

%
\setcounter{figure}{1}
\begin{figure}
\centering
\begin{tabular}{@{}c@{}}

\includegraphics{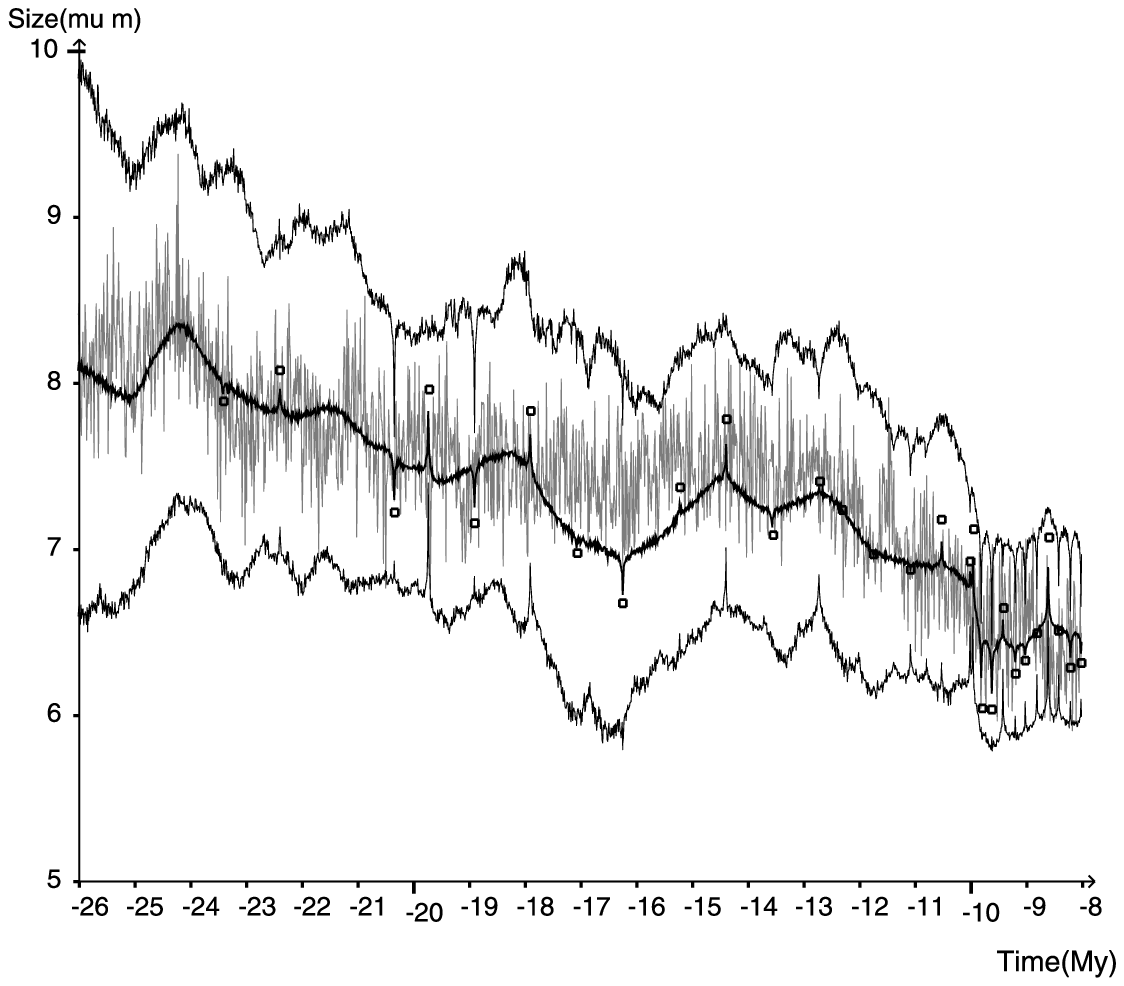}
\\
\footnotesize{(b)}
\end{tabular}
\caption{(Continued).}
\end{figure}

Looking at property inference, as described in
Section~\ref{modelprob}, we get results which are shown in
Table~\ref{properties}.
The data give little posterior weight to a process having only one
layer, while higher weights are given to two or three layers.
Also, there is an OU process rather than a random walk
in the lowest layer (thus stationarity) and there is evidence for
inter-regional correlations.
Three layers seem to be slightly preferred over two layers and
regional pull parameters are found in the middle layer.
There seems to be support for the existence of at least one
deterministic layer.
These results support models having the same overall
structure as the top model described in Table~\ref{bestmodnoid}.
All of the top five models under both AIC and BML selection, described
in the supplementary material [\citet{supp}], also seem to have the same properties.

%
\begin{table}
\tabcolsep=0pt
\caption{Posterior weights for different properties, with the
number of models in parenthesis. Note that while regional parameters
can be found either in the first, second or third layer in
a~three-layered model, such a feature is only possible in the first and
second layer in a~two-layered model and only in the first layer in
a~one-layered model. Similarly, a~three-layered model may be allowed to
have no diffusion either in the first or second layer or both, but
a~two-layered model may only be without diffusion in the first layer and
a layer without diffusion is not an option for a one-layered model}
\label{properties}
\begin{tabular*}{\textwidth}{@{\extracolsep{\fill}}lcccc@{}}
\hline
\textbf{Property} & \multicolumn{4}{c@{}}{\textbf{Options}} \\
\hline
Number of layers: & 1 & 2 & 3 & \\
& 7.5\% (18) & 32.4\% (114) & 60.1\% (578) & \\[3pt]
Regionality in:& none or $\mu_0$ & pull & diffusion & \\
& 0.2\% (177) & 88.3\% (259) & 11.5\% (274) & \\[3pt]
Regional parameters in: & no layer & layer 1 & layer 2 & layer 3 \\
& 0.2\% (177) & 13.1\% (205) & 69.1\% (196) & 17.5\% (132) \\[3pt]
Inter-regional & none & intermediate (6D) & perfect (1D) & \\
\quad instantaneous
correlations: & 4.9\% (50) & 85.2\% (212) & 9.9\% (448) & \\ [3pt]
No diffusion in & no layer & layer 1 & layer 2 & both layer 1 and 2 \\
& 7.7\% (486) & 1.2\% (132) &
91.1\% (72) & 0.007\% (20) \\[3pt]
Random walk in & no & yes & & \\
\quad the lowest layer: & 99.1\% (414) & 0.9\% (296) & & \\
\hline
\end{tabular*}
\end{table}

Bayesian model comparison can be hampered by sensitivity to
the prior probabilities. We therefore did a new analysis with
another prior distribution deemed reasonable, namely,
one where the characteristic times were given 95\%
credibility bands spanning from $1$~y to $300$~My rather than
$1$~ky to $1$~Gy and where the distribution of the diffusion
parameters were widened. The top two models according to BML
remained the same and the parameter estimates remained
essentially the same also. The property weights were subtly
shifted, but not enough to change any previous conclusions.
These results suggest to us that the Bayesian inference was
robust. See the supplementary material [\citet{supp}] for details concerning
various sensitivity tests.

The structure of the inter-regional correlations can be
investigated by categorizing according to in which layer
the inter-regional correlations exist. The posterior weights
(Table~\ref{corrprop}) indicate support for
inter-regional correlations in the lowest layer and possibly also on
the top layer. According to the best model in
Table~\ref{bestmodnoid}, the top layer is found to be
very fast moving (having a characteristic time less than
$10$~ky, which is less than the smallest time intervals in the data).
With the rather coarse time resolution in our data, the issue
of inter-regional correlations in the upper layer must therefore
remain unresolved.

%
\begin{table}
\caption{Posterior weights of correlation structure properties.
Only three-layered models with
nondegenerate pull at the bottom layer (333 in total) were examined}
\label{corrprop}
\begin{tabular*}{\textwidth}{@{\extracolsep{\fill}}ld{2.1}d{2.1}d{2.1}d{2.1}d{2.1}d{2.1}d{2.1}d{2.1}@{}}
\hline
\multicolumn{9}{@{}l}{\textbf{Inter-regional correlation in}} \\
\hline
Top layer & \multicolumn{1}{c}{N} & \multicolumn{1}{c}{Y} & \multicolumn{1}{c}{N} & \multicolumn{1}{c}{Y} &
\multicolumn{1}{c}{N} & \multicolumn{1}{c}{Y} & \multicolumn{1}{c}{N} & \multicolumn{1}{c}{Y} \\
Intermediate layer & \multicolumn{1}{c}{N} & \multicolumn{1}{c}{N} & \multicolumn{1}{c}{Y} &
\multicolumn{1}{c}{Y} & \multicolumn{1}{c}{N} & \multicolumn{1}{c}{N} & \multicolumn{1}{c}{Y} & \multicolumn{1}{c}{Y} \\
Bottom layer & \multicolumn{1}{c}{N} & \multicolumn{1}{c}{N} & \multicolumn{1}{c}{N} & \multicolumn{1}{c}{N} & \multicolumn{1}{c}{Y} &
 \multicolumn{1}{c}{Y} & \multicolumn{1}{c}{Y} & \multicolumn{1}{c}{Y} \\
Number of models & 15 & 16 & 30 & 32 & 56 & 60 & 60 & 64 \\
Posterior probability (\%) & 1.6 & 1.1 &
4.8 & 2.5 & 49.9 & 25.2 &
10.1 & 4.9 \\
\hline
\end{tabular*}
\end{table}

Although our study is explorative, we are confident that \textit{Coccolithus}
has phenotypic evolution with motion in at least
two layers. In addition, the
posterior distributions for the pull parameter(s) in the
second layer indicate that the fitness
optima are not evolving as random walks, which was also the case
for the optimum in \citet{hansen2008}, though there the optimum
was modeled to track a random walk in the lower layer, while in
our study the data suggests a stationary lower layer.

\section{Summary}
\label{conclusions}
A modeling framework for systems of related processes evolving in
continuous time has been constructed. This framework has been
applied to fossil data of size variability in marine unicellular
algae spanning nearly 60 million years. A simulation study showed
that for the amount of data in this application, the number of layers
could be inferred.

There is basic consensus among the models considered best using
different criteria and in the property analysis as to the
following model properties:

(1) There is more than one layer of stochastic processes
at play.
(2) There is dependency between what happens in two different regions.
(3) The are some regional differences in the nature of the dynamics,
that is, there are some parameters that are site-specific.
(4) The regional dependency does not stem primarily from
the top layer, but from something further down the causal chain.
(5) The hidden process seems to exhibit slow variation resulting in
long autocorrelation in cell size.

This suggests that both global and local
processes influence the mean phenotype, through dynamic
site-specific fitness optima that respond to an underlying
process that is partly global. However, as already mentioned,
we find no support for a forcing by a global temperature
indicator series on \textit{Coccolithus} size, in
contrast to reports on other biotic groups
[see \citet{schmidt2004}, \citet{finkel2005}].
The regional differences in estimated model parameters might
reflect contrasts in local environmental conditions, differences
in morphotype [possibly (sub)species] composition or a
combination of both. These interpretations and
additional sensitivity tests (e.g., how age models affect the
model outcome) will be further explored in another publication
[see \citet{henderiks2011}].\vadjust{\goodbreak}

Thus, it seems that the framework can be used for studying and reaching
tentative conclusions about the driving forces behind a phenotypic
time series. It should also be possible to use this framework in
other settings both within and outside paleontology and
evolutionary biology. Time series with irregular temporal resolution,
due to missing data, breaks in the observational scheme or for
other reasons, are not uncommon. For such data our framework provides
an alternative to methods based on auto regression with regular
temporal resolution.

The class of models described by linear SDEs is wide, but
computationally feasible. They allow causality to be modeled and
studied [see \citet{schweder2012}], and they accommodate
latent hierarchical structures.
Our model could, for instance, be expanded by including more
layers, allowing for time series that are phylogenetically
related, allowing for external processes being included as
exogenous forcings in different layers or by
using the framework also for modeling these time-series,
thus making them an integral part of the analysis.
Furthermore, geographic information could be incorporated by letting
the correlation between sites depend on the distance between
the sites, for example, in relation to ocean circulation.

\section*{Source code}

The source codes for our analysis programs can be found on
the web page \url{http://folk.uio.no/trondr/layered}.

\section*{Acknowledgments}

We would like to thank Thomas Hansen for helpful comments.
We also thank the reviewers and the editor for their extensive advice.

\begin{supplement}
\stitle{Phenotypic evolution studied by layered stochastic
differential equations---supplementary material}
\slink[doi]{10.1214/12-AOAS559SUPP} 
\slink[url]{http://lib.stat.cmu.edu/aoas/559/supplement.pdf}
\sdatatype{.pdf}
\sdescription{Supplementary material: Mathematical details, description of the prior distribution,
Kalman filtering, practical restrictions, numerical methods,
data issues, extra material on simulation studies and model selection results, and robustness analysis.}
\end{supplement}

%

\printaddresses

\end{document}